\documentclass[12pt,thmsa]{article}
\usepackage{amssymb}
\usepackage{makeidx}
\usepackage{amsfonts}
\usepackage{amsmath}
\usepackage{sw20lart}

\input tcilatex
\begin{document}

\author{Nick Laskin\thanks{\textit{E-mail address}: nlaskin@rocketmail.com}}
\title{\textbf{Quantum Well in Fractional Quantum Mechanics}}
\date{TopQuark Inc.\\
Toronto, ON, M6P 2P2\\
Canada}
\maketitle

\begin{abstract}
Within the framework of fractional quantum mechanics, an exact solution has
been found for the energy spectrum of a quantum particle confined in a
quantum well - a symmetric one-dimensional finite potential well. A simple
graphical algorithm is described for obtaining the number of discrete levels
in a quantum well and their associated energy values.

The presented results open up new possibilities for emulating fractional
quantum mechanics using quantum wells.


\end{abstract}

\section{Introduction}

A simple and exactly solvable problem of quantum mechanics is the problem of
finding the spectrum of a particle confined in a one-dimensional (1D)
potential field having a rectangular shape, see \S 22 in \cite%
{LandauLifshitzQM}. Such a potential field is called a rectangular potential
well. A potential well can have infinite or finite depth.

Interest in 1D quantum mechanical models is driven by the study, design and
developments of quantum wells. Quantum well (QW) is a semiconductor layered
structure that fabricated by sandwiching a thin layer of semiconductor "well
material" between two layers of a semiconductor "barrier material".
Comparing to well material the barrier material has a wider band gap. As an
example of such layered semiconductors structure let's mention layer of
gallium arsenide GaAs (well material) sandwiched between two layers of
aluminum gallium arsenide AlGaAs (barrier material). Another example is a
thin layer of indium gallium nitride InGaN (well material) sandwiched
between two layers of gallium nitride GaN (barrier material). Due to the
difference in band gaps between InGaN and GaN, there is a 1D quantum
potential well that confines electrons and holes, resulting in quantum
confinement. As a result of the confinement, the energy levels of the
confined particles are quantized, that is, they become discrete. The number
of discrete levels in QW and their corresponding energies can be tuned by
adjusting the depth and thickness of QW. Which in turn allows the emission
of light at specific wavelengths to be controlled. GaN-based devices such as
light-emitting diodes (LEDs), laser diodes (LDs), and photodetectors are
widely used in many fields.

It is well known that impurities and their interactions with electrons and
holes affect the electronic and optical properties of semiconductor
materials. The impurities are distributed randomly in a semiconductor
material. Therefore, the problem of modeling the energy spectrum in QW
requires the use of the Schr\"{o}dinger equation with a potential modified
by stochastic component induced by impurities. In \cite{Stephanovich1} it
was suggested to mock-up the disorder in QW implementing fractional Schr\"{o}%
dinger equation with regular (non stochastic) potential. The fractality
parameter $\alpha $ $(1<\alpha \leq 2)$, the Levy index \cite{LaskinFQM},
allows modeling the impact of the impurities distribution in a semiconductor
on the energy spectrum of a quantum particle in QW. As shown in the work 
\cite{Stephanovich1}, "weak disorder" is characterized by a Gaussian
probability distribution, which is local and occurs when the parameter $%
\alpha =2$. The case of "strong disorder" occurs when $\alpha $ deviates
from 2, leading to long-range non-local distributions. Thus, the idea is to
use framework of fractional quantum mechanics \cite{LaskinFQM} to model
energy spectrum in QW.

In the fractional quantum mechanics \cite{LaskinFQM} the problem of finding
the spectrum of a particle confined in 1D infinite potential well was solved
first in \cite{LaskinChaos}. Despite the initial controversy surrounding
this exact solution, it is now a generally accepted result thanks to S.
Bayin \cite{Bayin}, who proved the correctness of the solution obtained in 
\cite{LaskinChaos} by directly substituting this solution into the 1D
fractional Schr\"{o}dinger equation and performing the calculations.

It turns out that in fractional quantum mechanics the problem of finding the
spectrum of quantum mechanical particle moving in symmetrical 1D finite
potential well is also exactly solvable. Moreover, the problem is reduced to
finding the roots of a transcendental equations, which are exactly the same
as in standard quantum mechanics. The only difference between fractional and
standard quantum mechanics is the dispersion law, that is, the fundamental
relationship between the energy and momentum of a quantum particle.

This paper presents exact solutions for the energy spectrum of a particle
confined in symmetric 1D rectangular potential well in the framework of
fractional quantum mechanics.

The paper is organized as follows. Section 2 presents the 1D fractional Schr%
\"{o}dinger equation and its general solutions for a particle moving in the
field of a one-dimensional finite symmetric potential well. Section 3
discusses in detail the boundary conditions at the edges of the potential
well in order to obtain transcendental equations for finding the energy
spectra of even and odd quantum states of a particle in a quantum well.
Section 4 introduces dimensionless variables to simplify the transcendental
equations and describes a simple graphical algorithm for solving the
resulting system of transcendental equations to find number of energy levels
in the well and the energy values associated with them.

In Conclusion, the obtained results are summarized and discussed.

\section{Fractional Schr\"{o}dinger equation for a particle moving in 1D
finite symmetric potential well}

The 1D fractional Schr\"{o}dinger equation first introduced by Laskin has
the form \cite{LaskinFQM}, \cite{Laskin4}, \cite{LaskinfSch}

\begin{equation}
i\hbar \frac{\partial \psi (x,t)}{\partial t}=-D_{\alpha }(\hbar \nabla
)^{\alpha }\psi (x,t)+V(x)\psi (x,t),\qquad 1<\alpha \leq 2,  \label{eq10sh}
\end{equation}%
where $\psi (x,t)$ is the wave function, $\hbar $ is Planck's constant, $%
D_{\alpha }$ is the scale factor with units of $[D_{\alpha }]=\mathrm{erg}%
^{1-\alpha }\mathrm{cm}^{\alpha }\sec ^{-\alpha }$\ (when $\alpha =2$, $%
D_{2}=1/2m$, where $m$ is the mass of the quantum particle), and the 1D
quantum Riesz fractional derivative $(\hbar \nabla )^{\alpha }$ was defined
as follows \cite{LaskinFQM}, \cite{Laskin4}

\begin{equation}
(\hbar \nabla )^{\alpha }\psi (x,t)=-\frac{1}{2\pi \hbar }%
\int\limits_{-\infty }^{\infty }dpe^{i\frac{px}{\hbar }}|p|^{\alpha }\varphi
(p,t),  \label{eq11R}
\end{equation}%
here $\varphi (p,t)$ is the Fourier transform of the wave function $\psi
(x,t)$

\begin{equation}
\varphi (p,t)=\int\limits_{-\infty }^{\infty }dxe^{-i\frac{px}{\hbar }}\psi
(x,t),  \label{eq2F}
\end{equation}%
and respectively

\begin{equation}
\psi (x,t)=\frac{1}{2\pi \hbar }\int\limits_{-\infty }^{\infty }dpe^{i\frac{%
px}{\hbar }}\varphi (p,t).  \label{eq3F}
\end{equation}

Let us consider the problem of finding energy levels for a quantum particle
moving in a symmetric potential field $V(x)$ having a rectangular shape

\begin{equation*}
V(x)=U,\qquad x<-a,\qquad \qquad (\mathrm{i}),
\end{equation*}

\begin{equation}
V(x)=0,\quad -a\leq x\leq a,\quad \qquad (\mathrm{ii}),  \label{eq1}
\end{equation}

\begin{equation*}
V(x)=U,\qquad \ x>a.\qquad \qquad \ (\mathrm{iii}).
\end{equation*}

We are interested in solutions of the fractional Schr\"{o}dinger equation (%
\ref{eq10sh}) for stationary states when $E<U$. A stationary state with an
energy $E$ is described by a wave function $\psi (x,t)$, which can be
written as $\psi (x,t)=\exp \{-iEt/\hbar \}\phi (x),$ where $\phi (x)$ is
time independent wave function that satisfies the time independent
fractional Schr\"{o}dinger equation

\begin{equation}
-D_{\alpha }(\hbar \nabla )^{\alpha }\phi (x)+V(x)\phi (x)=E\phi (x).
\label{eq136}
\end{equation}

Outside of the potential well, regions $(\mathrm{i})$ and $(\mathrm{iii}),$
Eq.(\ref{eq136}) takes the form

\begin{equation}
-D_{\alpha }(\hbar \nabla )^{\alpha }\phi _{\mathrm{i,iii}}(x)+U\phi _{%
\mathrm{i,iii}}(x)=E\phi _{\mathrm{i,iii}}(x),  \label{eq136out}
\end{equation}

and inside of the well, in the region $(\mathrm{i}),$ it takes the form

\begin{equation}
-D_{\alpha }(\hbar \nabla )^{\alpha }\phi _{\mathrm{ii}}(x)=E\phi _{\mathrm{%
ii}}(x),  \label{eq136in}
\end{equation}

where $\phi _{\mathrm{i}}(x),$ $\phi _{\mathrm{ii}}(x)$ and $\phi _{\mathrm{%
iii}}(x)$ are the wave functions in the regions $(\mathrm{i}),(\mathrm{ii})$
and $(\mathrm{iii})$ respectively.

Using the property of the quantum fractional Riesz derivative given by
Eq.(3.42) in \cite{LaskinFQM}, we express the general solutions of the Eq.(%
\ref{eq136out}) for the wave function outside the well $(|x|>a)$ in the form

\begin{equation}
\phi _{\mathrm{i,iii}}(x)=Ae^{\kappa x}+Be^{-\kappa x},  \label{eq137sol}
\end{equation}

where $A$ and $B$ are normalization constants and 

\begin{equation}
\kappa =\frac{1}{\hbar }(\frac{U-E}{D_{\alpha }})^{1/\alpha },\quad
E<U,\qquad 1<\alpha \leq 2,  \label{eq137kapa}
\end{equation}

and inside the well $(|x|<a)$ the general solution of Eq.(\ref{eq136in}) has
the form

\begin{equation}
\phi _{\mathrm{ii}}(x)=C\cos \mathrm{k}x+D\sin \mathrm{k}x,  \label{eq138sol}
\end{equation}

where $C$ and $D$ are normalization constants and 

\begin{equation}
\mathrm{k}=\frac{1}{\hbar }(\frac{E}{D_{\alpha }})^{1/\alpha },\qquad
1<\alpha \leq 2.  \label{eq138k}
\end{equation}

\section{Boundary conditions and energy spectrum}

Imposing a boundary condition at $x=\pm a$ on the wave function allows one
to obtain transcendental equations for finding the energy spectrum of a
quantum particle in a finite symmetric potential well. From the requirement
of continuity of the wave function at the edges of the well, we obtain the
following boundary conditions:

\begin{equation}
\phi _{\mathrm{i}}(-a)=\phi _{\mathrm{ii}}(-a),\qquad \phi _{\mathrm{ii}%
}(a)=\phi _{\mathrm{iii}}(a),  \label{eq136bound}
\end{equation}

and

\begin{equation}
\phi _{\mathrm{i}}^{\prime }(-a)=\phi _{\mathrm{ii}}^{^{\prime }}(-a),\qquad
\phi _{\mathrm{ii}}^{^{\prime }}(a)=\phi _{\mathrm{iii}}^{^{\prime }}(a),
\label{eq137bound}
\end{equation}

here the following notations were introduced

\begin{equation*}
\phi _{\mathrm{i,ii,iii}}^{\prime }(\pm a)=\frac{d}{dx}\phi _{\mathrm{%
i,ii,iii}}(x)|_{x=\pm a}.
\end{equation*}

Due to the symmetry of the well potential $V(x)=V(-x)$, the solution of the
fractional Schr\"{o}dinger equation (\ref{eq10sh}) will be inside the well
either symmetric (even) or antisymmetric (odd) relative to the well center.
Considering the even solution $\phi (x)=\phi (-x)$, we come to the
conclusion that $D=0$ in Eq.(\ref{eq138sol}). Thus, the even solution for
the wave function inside the well has the form

\begin{equation}
\phi _{\mathrm{ii}}^{(\mathrm{even})}(x)=C\cos \mathrm{k}x.  \label{eq139in}
\end{equation}

Outside the potential well the wave function decays exponentially

\begin{equation}
\phi _{\mathrm{i}}^{(\mathrm{even})}(x)=Be^{-\kappa x},\qquad x>a,
\label{eq139out}
\end{equation}

and

\begin{equation*}
\phi _{\mathrm{iii}}^{(\mathrm{even})}(x)=Ae^{\kappa x},\qquad x<-a.
\end{equation*}

Imposing the boundary condition at $x=a$,\ we have

\begin{equation}
C\cos \mathrm{k}a=Be^{-\kappa a},  \label{eq41a}
\end{equation}

and%
\begin{equation}
\mathrm{k}C\sin \mathrm{k}a=\kappa Be^{-\kappa a}.  \label{eq41b}
\end{equation}

Imposing the boundary condition at $x=-a$,\ we have

\begin{equation}
C\cos \mathrm{k}(\mathrm{-}a)=Ae^{-\kappa a},  \label{eq41c}
\end{equation}

and%
\begin{equation}
-\mathrm{k}C\sin \mathrm{k}(-a)=\kappa Ae^{-\kappa a}.  \label{eq41d}
\end{equation}

Dividing either Eq.(\ref{eq41b}) by Eq.(\ref{eq41a}) or Eq.(\ref{eq41d}) by
Eq.(\ref{eq41c}) we obtain

\begin{equation}
\kappa =\mathrm{k\tan k}a,\qquad (\mathrm{even}).  \label{eq42b}
\end{equation}

By solving this transcendental equation, we find energy spectrum of a
quantum particle in the finite symmetric potential well in the case when the
wave function of the particle is symmetric (even), that is $\phi (x)=\phi
(-x)$. Note that Eq.(\ref{eq42b}) is valid for fractional quantum mechanics
when $1<\alpha <2$, and also for standard quantum mechanics when $\alpha =2$%
. In the case of standard quantum mechanics Eqs.(\ref{eq137kapa}) and (\ref%
{eq138k})\ read \cite{LandauLifshitzQM}

\begin{equation}
\kappa =\frac{1}{\hbar }\sqrt{2m(U-E)},\quad E<U,  \label{eq137st}
\end{equation}

and 

\begin{equation}
\mathrm{k}=\frac{1}{\hbar }\sqrt{2mE)},  \label{eq138st}
\end{equation}

where $m$ is the mass of the quantum particle.

Considering the odd solution $\phi (x)=-\phi (-x)$, we come to the
conclusion that $C=0$ in Eq.(\ref{eq138sol}). Thus, the odd solution for the
wave function inside the well has the form

\begin{equation}
\phi _{\mathrm{ii}}^{(\mathrm{odd})}(x)=D\sin \mathrm{k}x.  \label{eq140in}
\end{equation}

Outside the potential well the wave function decays exponentially

\begin{equation}
\phi _{\mathrm{i}}^{(\mathrm{odd})}(x)=Be^{-\kappa x},\qquad x>a,
\label{eq140out}
\end{equation}

\begin{equation*}
\phi _{\mathrm{iii}}^{(\mathrm{odd})}(x)=Ae^{\kappa x},\qquad x<-a.
\end{equation*}

Imposing the boundary condition at $x=a$,\ we have

\begin{equation}
D\sin \mathrm{k}a=-Be^{-\kappa a},  \label{eq43a}
\end{equation}

and%
\begin{equation}
-\mathrm{k}D\cos \mathrm{k}a=-\kappa Be^{-\kappa a}.  \label{eq43b}
\end{equation}

Imposing the boundary condition at $x=-a$,\ we have

\begin{equation}
D\sin \mathrm{k}(\mathrm{-}a)=Ae^{-\kappa a},  \label{eq43c}
\end{equation}

and%
\begin{equation}
-\mathrm{k}D\cos \mathrm{k}(-a)=\kappa Ae^{-\kappa a}.  \label{eq43d}
\end{equation}

Dividing either Eq.(\ref{eq43b}) by Eq.(\ref{eq43a}) or Eq.(\ref{eq43d}) by
Eq.(\ref{eq43c}) we obtain

\begin{equation}
\kappa =-\mathrm{k\cot k}a,\qquad (\mathrm{odd}).  \label{eq44}
\end{equation}

By solving this transcendental equation, we find energy spectrum of a
quantum particle in a finite symmetric potential well in the case when the
wave function of the particle is antisymmetric (odd), that is $\phi
(x)=-\phi (-x)$. Note that Eq.(\ref{eq44}) is valid for fractional quantum
mechanics when $1<\alpha <2$, and also for standard quantum mechanics when $%
\alpha =2$. In the case of standard quantum mechanics, $\kappa $ and $%
\mathrm{k}$ are given by Eqs.(\ref{eq137st}) and (\ref{eq138st}),
respectively see \S 22 in \cite{LandauLifshitzQM}.

\section{Dimensionless variables}

Instead of $\kappa $ and $\mathrm{k}$ defined by Eqs.(\ref{eq137kapa}) and (%
\ref{eq138k})\ we introduce dimensionless variables $\eta $ and $\varsigma $

\begin{equation}
\eta =\kappa a=\frac{a}{\hbar }(\frac{U-E}{D_{\alpha }})^{1/\alpha },\quad
E<U,\qquad 1<\alpha \leq 2,  \label{eqD2}
\end{equation}

and

\begin{equation}
\varsigma =\mathrm{k}a=\frac{a}{\hbar }(\frac{E}{D_{\alpha }})^{1/\alpha
},\qquad 1<\alpha \leq 2,  \label{eqD1}
\end{equation}

respectively. In terms of $\eta $ and $\varsigma $ Eqs.(\ref{eq42b}) and (%
\ref{eq44}) become

\begin{equation}
\eta =\varsigma \tan \varsigma ,\qquad \eta ,\,\varsigma \geq 0,\qquad (%
\mathrm{even}),  \label{eqD4}
\end{equation}

and

\begin{equation}
\eta =-\varsigma \cot \varsigma ,\qquad \eta ,\,\varsigma \geq 0,\qquad (%
\mathrm{odd}).  \label{eqD5}
\end{equation}

It is easy to see that

\begin{equation}
\eta ^{\alpha }+\varsigma ^{\alpha }=(\kappa a)^{\alpha }+(\mathrm{k}%
a)^{\alpha }=\frac{a^{\alpha }U}{\hbar ^{\alpha }D_{\alpha }}=\mathrm{const}%
,\qquad 1<\alpha \leq 2,  \label{eqD3}
\end{equation}

where $\mathrm{const}$ depends on the well depth $U$, its width $a$, Plank's
constant $\hbar $, the scale factor $D_{\alpha }$ and the fractality
parameter $\alpha $.

Thus, a simple graphical algorithm for solving the system of equations (\ref%
{eqD4}) - (\ref{eqD3}) consists of choosing the numerical value of the
dimensionless parameter $(a^{\alpha }U)/(\hbar ^{\alpha }D_{\alpha })$,
plotting three curves and collecting data on the intersection points. The
intersection points of the curves (\ref{eqD4}) - (\ref{eqD3}) give the
values for $\varsigma $ and $\eta $. The number of the intersection points
depends on value of the dimensionless parameter $(a^{\alpha }U)/(\hbar
^{\alpha }D_{\alpha })$. The value for $\varsigma $ from the intersection
points of curves (\ref{eqD4}) and (\ref{eqD3}) corresponds via Eq.(\ref{eqD1}%
) to a certain energy levels in the spectrum for an even quantum stationary
state for a particle confined in a 1D finite symmetric potential well. The
value of $\eta $ from the intersection points of curves (\ref{eqD5}) and (%
\ref{eqD3}) corresponds via Eq.(\ref{eqD2}) to a certain energy level in the
spectrum for an odd quantum stationary state for a particle confined in a 1D
finite symmetric potential well.

Thus, the solutions of the system of equations (\ref{eqD4}) - (\ref{eqD3})
determine the number of energy levels and the corresponding energy values
for even and odd quantum stationary states for a particle in 1D finite
symmetric potential well. 

The wave function of a quantum particle in a well oscillates inside the well
and decays exponentially outside it.

\section{Conclusion}

Within the framework of fractional quantum mechanics, an exact analytical
solution has been found for the energy spectrum of a quantum particle
confined in a quantum well - a symmetric one-dimensional finite potential
well. A simple numerical algorithm for obtaining the number of discrete
levels in a well and their associated energy values is proposed and
discussed. The obtained solutions are directly applicable for the study and
modeling of discrete quantum-mechanical states in sandwich structures of
semiconductors, for example, InGaN - GaN - InGaN.

The presented results open up new possibilities for emulating fractional
quantum mechanics using quantum wells.


\begin{thebibliography}{9}
\bibitem{LandauLifshitzQM} L. D. Landau and E. M. Lifshitz, \textit{Quantum
Mechanics - Nonrelativistic Theory,} Third Edition. Volume 3, Course of
Theoretical Physics, Pergamon Press, Oxford, 1976.

\bibitem{Stephanovich1} V. A. Stephanovich, E. V. Kirichenko, B. Pytel and
A. W\'{o}jcik, V. K. Dugaev and M. Inglot, The impact of disorder on the
charge carrier and exciton spectra in semiconductor quantum well structures,
https://papers.ssrn.com/sol3/papers.cfm?abstract\_id=5216352.

\bibitem{LaskinFQM} N. Laskin, \textit{Fractional Quantum Mechanics}, World
Scientific, 2018.

\bibitem{LaskinChaos} N. Laskin, Fractals and quantum mechanics, Chaos, 
\textbf{10}, (2000), 780--790.

\bibitem{Bayin} S. Bayin, On the consistency of the solutions of the space
fractional Schr\"{o}dinger equation, J. Math. Phys. \textbf{53}, (2012),
042105. (Available on-line at https://arxiv.org/pdf/1203.4556).

\bibitem{Laskin4} N. Laskin, Fractional quantum mechanics and L\'{e}vy path
integrals, Phys. Lett., \textbf{A268}, (2000), 298-305.

\bibitem{LaskinfSch} N. Laskin, Fractional Schr\"{o}dinger equation, Phys.
Rev. \textbf{E66}, (2002), 056108.
\end{thebibliography}
\end{document}